\documentclass[runningheads,orivec]{llncs}

\usepackage{hyphenat}
\usepackage{fix-cm}
\usepackage[T1]{fontenc}
\usepackage{todonotes}
\usepackage{booktabs}
\usepackage{diagbox}
\usepackage{enumitem}
\usepackage{tikz}
\usepackage{xfp}
\usepackage{pifont} 
\newcommand*\circled[1]{\small\tikz[baseline=(char.base)]{
            \node[shape=circle,draw,inner sep=0.8pt] (char) {#1};}}
\usepackage{microtype}
\usepackage{multirow}
\usepackage{siunitx}
\usepackage{array}
\usepackage{amssymb}
\usepackage{graphicx}
\usepackage[hidelinks]{hyperref}
\usepackage[table,dvipsnames]{xcolor}
\usepackage[most]{tcolorbox}
\usepackage{color}
\usepackage{colortbl} 
\usepackage{cite}
\usepackage{pgf}
\usepackage{collcell}
\usepackage{fontawesome5}

\newcommand{\github}[1]{\faIcon{github}~\url{#1}}

\definecolor{st}{HTML}{D1E091}
\definecolor{rd}{HTML}{F3DB87}
\definecolor{wst}{HTML}{FF948A}
\colorlet{colorFst}{st}       
\colorlet{colorSnd}{st!60} 
\colorlet{colorTrd}{rd!50}      
\colorlet{colorWrst}{wst!40}
\newcommand{\fst}{\cellcolor{colorFst}\bf}   
\newcommand{\nd}{\cellcolor{colorSnd}}      
\newcommand{\rd}{\cellcolor{colorTrd}}      
\newcommand{\wst}{\cellcolor{colorWrst}} 

\newcommand\answerbox[2]{
\smallskip
    \begin{tcolorbox}[left=6pt, right=6pt, top=6pt, bottom=6pt,colback=st!30,colframe=st,coltext=black,title={#1},boxrule=0.5pt,width=\linewidth,coltitle=black,breakable,lefttitle=6pt,righttitle=6pt,toptitle=3pt,bottomtitle=3pt]
    {#2}
    \end{tcolorbox}
}

\newcommand{\better}{\textcolor{Green}{\(\uparrow\)\hspace{0.2em}}}    
\newcommand{\worse}{\textcolor{red}{\(\downarrow\)\hspace{0.2em}}}   

\newcommand{\orcidicon}{\includegraphics[width=0.32cm]{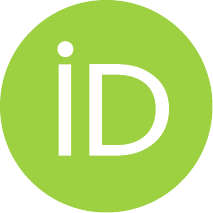}}
\foreach \x in {A, ..., Z}{%
\expandafter\xdef\csname orcid\x\endcsname{\noexpand\href{https://orcid.org/\csname orcidauthor\x\endcsname}{\noexpand\orcidicon}}
}

\begin{document}
\title{Revisiting Vul-RAG: Reproducibility and Replicability of RAG-based Vulnerability Detection with Open-Weight Models}
\titlerunning{Revisiting Vul-RAG}

\author{Sabrina Kaniewski\inst{1} \and Fabian Schmidt\inst{2} \and
Tobias Heer\inst{1}}

\authorrunning{S. Kaniewski et al.}
\institute{Institute for Secure Networked Systems, Esslingen University, Esslingen, Germany \and
Institute for Intelligent Systems, Esslingen University, Esslingen, Germany \\ Contact: \email{sabrina.kaniewski@hs-esslingen.de}}


%
\maketitle
\begin{abstract}
Large language models (LLMs) have shown strong potential for automated software vulnerability detection, particularly in retrieval-augmented generation (RAG) settings. However, for approaches relying on proprietary models and APIs, reproducibility and replicability remain largely unexplored, raising the question of whether reported results generalize or depend primarily on specific model choices. In this work, we present a reproducibility study of Vul-RAG, a RAG-based framework for source code vulnerability detection that enhances LLMs with high-level vulnerability knowledge. We first replicate the results in a fully local and open-weights setting using the reported open-weight baseline models. We then extend the evaluation to a diverse set of recent open-weight LLMs, including code-specialized, general-purpose, and reasoning models of varying parameter sizes. The results confirm that the findings of Vul-RAG are reproducible under local deployment, but with minor deviations. Across all evaluated models, we observe a performance plateau at approximately 0.30 pairwise accuracy (code pairs for which both the vulnerable and the patched function are correctly classified). Notably, this plateau persists even for more recent and advanced models, indicating that improvements in model capacity alone do not substantially enhance performance. Finally, we discuss practical implications and trade-offs between detection effectiveness, model capabilities, and model scale. Implementation and evaluation artifacts are publicly available at \url{https://github.com/hs-esslingen-it-security/revisiting-Vul-RAG}.
\end{abstract}


%
\section{Introduction}
Software vulnerability detection is a critical task in the software development life cycle, enabling developers to identify and mitigate security issues before they can be exploited in deployed systems~\cite{kaniewski2025handling}.  
Recent advances in Large Language Models (LLMs) have opened new opportunities for automating vulnerability detection. 
Due to their extensive world knowledge and ability to generalize across programming languages and code representations, LLMs have emerged as a promising approach for analyzing the security of source code~\cite{kaniewski2025systematicliteraturereviewdetecting}. 

One particularly promising technique for adapting LLMs to the software vulnerability detection task is Retrieval-Augmented Generation (RAG)~\cite{lewis2020retrieval}. RAG enhances LLMs by dynamically retrieving relevant information from a knowledge base and incorporating it into the prompt during inference. By augmenting prompts with up-to-date vulnerability knowledge, RAG allows LLMs to reason about vulnerabilities beyond the information contained in the model’s training data~\cite{kaniewski2025systematicliteraturereviewdetecting}.
Recent studies explore different knowledge representations such as vulnerability descriptions or labeled code examples to improve detection performance~\cite{antalEvaluatingRetrievalAugmentedGeneration2026,du2025vulragenhancingllmbasedvulnerability,luGRACEEmpoweringLLMbased2024,safdarRealVulLLMLLMBased2025,sunLLM4VulnUnifiedEvaluation2024,tsaiLeveragingIntraInterReferences2025,tsaiSequentialMultiStageApproach2025,wenVulEvalRepositoryLevelEvaluation2024,wuMulVulRetrievalaugmentedMultiAgent2026,zhuSpecificationGuidedVulnerabilityDetection2025}.

Despite advancements in LLM-based software vulnerability detection, several challenges hinder performance validation.
Many studies do not release their code and artifacts as open-source or rely on proprietary models and APIs such as OpenAI's GPT. 
It remains unclear whether reported performance gains are a result of the proposed methodologies or the selection of highly capable, proprietary models.
Reproducibility, i.e., the ability for an independent group of researchers to obtain the same results using the provided methods and data, is a fundamental requirement for validating findings and incremental knowledge building~\cite{angermeir2025reflectionsreproducibility}. 
In LLM-based research, reproducibility is further complicated by rapidly evolving model ecosystems, undocumented dependency changes, and the non-deterministic nature of outputs~\cite{sallouBreakingSilenceThreats2024}. 
In addition, the reliance on external APIs raises concerns for security-critical domains such as vulnerability detection, where sensitive or proprietary code may not be suited for transmission to third-party services.

\textbf{Contributions}. 
To address these challenges, we conduct a reproducibility study of Vul-RAG~\cite{du2025vulragenhancingllmbasedvulnerability}, a recently proposed RAG-based framework for source code vulnerability detection.
The evaluation of Vul-RAG relied on a set of proprietary and open-weight models.
Given the rapid evolution of the model ecosystem, there is a need to evaluate whether these findings hold for the latest generation of models and to identify current limitations in cross-model replicability.
To enable reproducible experimentation, we adapt the Vul-RAG framework to operate entirely with open-weight models and local inference, removing the dependency on proprietary APIs.
First, we reproduce the results using the reported open-weight baseline models. 
Second, we systematically evaluate Vul-RAG using recent open-weight LLMs, including code-specialized, general-purpose, and reasoning models of various parameter sizes.
Through this study, we aim to assess the reproducibility of the reported results, investigate the sensitivity of Vul-RAG to the underlying model, and discuss implications for selecting and deploying models and RAG-based systems in detection pipelines. We publish all implementation and evaluation artifacts at \github{https://github.com/hs-esslingen-it-security/revisiting-Vul-RAG}.

The remainder of this work is structured as follows.
In Section~\ref{sec:related_work}, we review related studies on open science.
In Section~\ref{sec:vul_rag}, we provide an overview of the Vul-RAG framework.
Section~\ref{sec:experimental_setup} outlines the experimental setup, including the research questions, dataset, studied LLMs, metrics, and implementation details.
In Section~\ref{sec:results}, we discuss the results.
We address threats to validity (e.g., potential evaluation bias) in Section~\ref{sec:threats} and conclude this work in
Section~\ref{sec:conclusion}.


\section{Related Work}
\label{sec:related_work}
Despite the growing body of work on LLM-based software vulnerability detection~\cite{kaniewski2025systematicliteraturereviewdetecting}, only a limited number of studies examine the reproducibility or replicability of existing approaches. 

Nong et al.~\cite{nongopenscience} investigated open science practices in deep learning (DL)-based software vulnerability detection by analyzing 55 studies across four dimensions: availability, executability, reproducibility, and replicability. 
Only 14 studies provided publicly available artifacts.
Among these, three lacked sufficient documentation or complete implementations, rendering them non-executable.
To assess reproducibility, the authors measured the deviation between reported and reproduced F1 scores.
Artifacts were classified as reproducible, weakly reproducible, or not reproducible if the absolute deviation was less than 1\%, between 1\% and 5\%, or greater than 5\%, respectively.
Only 7 out of 8 executable artifacts were reproducible, and merely two were fully replicable, i.e., reported performance could be replicated against different training and testing datasets.
 
Chakraborty~\cite{chakraborty2024revisiting} revisited four DL-based vulnerability detection approaches.
When evaluated on the original datasets, the reproduced results largely matched the reported ones, typically differing by about one percentage point. 
However, when replicated on a real-world benchmark, the authors observed a substantial performance drop, highlighting limitations in their practical applicability.

Similarly, Steenhoek et al.~\cite{steenhoek2023empirical} reproduced nine state-of-the-art DL-based vulnerability detection approaches using two common benchmark datasets. 
For most models, reproduced results differed by less than 2\% from those reported.

Recent work has also highlighted reproducibility challenges in LLM-based software engineering research.
Angermeir et al.~\cite{angermeir2025reflectionsreproducibility} analyzed the reproducibility of 85 LLM-centric studies published in 2024. 
Among these, 69 studies relied on OpenAI services either as base models or for comparison.
Of the 18 studies that provided research artifacts, only five were complete and executable. 
However, none could be fully reproduced: two were partially reproducible, 
while three could not be reproduced at all. 
The authors identified several common obstacles to reproducibility, including non-obvious missing artifacts (e.g., individual missing scripts), unspecified dependency versions, code issues, and deprecation of LLMs.

Despite these efforts, no existing study has examined the reproducibility of LLM-based vulnerability detection approaches that employ advanced adaptation techniques such as RAG. 
Given the increasing adoption of RAG-based techniques in LLM-assisted software analysis, understanding their reproducibility is essential for ensuring reliable and verifiable research outcomes.
In this work, we address this gap by reproducing Vul-RAG~\cite{du2025vulragenhancingllmbasedvulnerability}, one of the most cited (at the time of writing) and openly available RAG-based frameworks for vulnerability detection.


\section{Vul-RAG: Knowledge-level RAG Framework}
\label{sec:vul_rag}
\begin{figure}[b!]
    \centering
    \includegraphics[width=\linewidth]{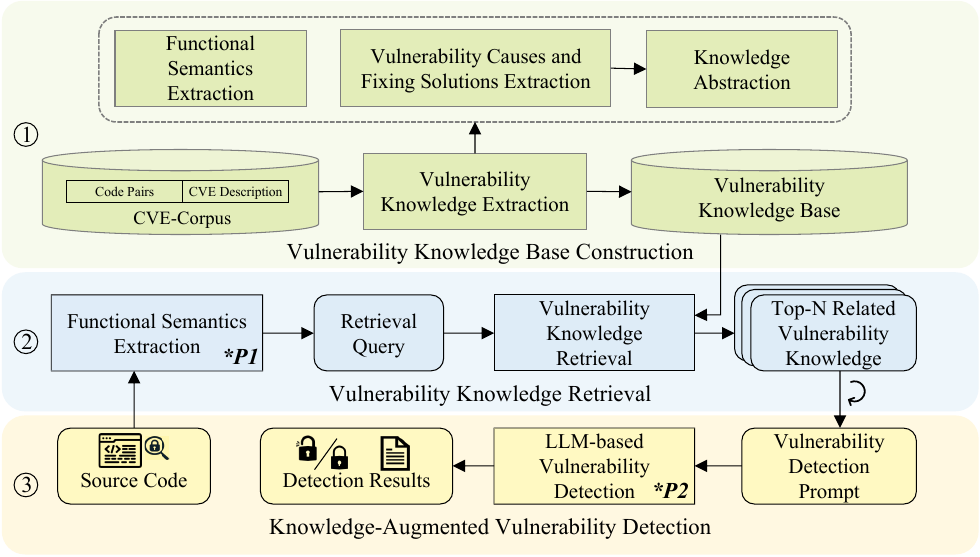}
    \caption{Overview of Vul-RAG, adapted from~\cite{du2025vulragenhancingllmbasedvulnerability}. The framework consists of three stages: \protect\circled{1}~offline vulnerability knowledge base construction; \protect\circled{2}~online retrieval of relevant knowledge items based on functional semantics; and \protect\circled{3}~iterative, knowledge-augmented vulnerability detection. The markers \textbf{\textit{*P1}} and \textbf{\textit{*P2}} highlight points of failure where model performance frequently failed, as discussed in Sections~\ref{sec:studied_llms} and \ref{sec:rq1}.}
    \label{fig:vul-rag}
    \vspace{-1.25em}
\end{figure}

Vul-RAG~\cite{du2025vulragenhancingllmbasedvulnerability} is a retrieval-augmented framework for source code vulnerability detection that enhances LLMs with high-level, generalizable vulnerability knowledge. 
Vul-RAG constructs a multi-dimensional representation of vulnerability knowledge that focuses on semantic and causal characteristics of vulnerabilities.
The framework consists of three steps as visualized in Figure~\ref{fig:vul-rag}:
\begin{enumerate}[label=\protect\circled{\arabic*}, leftmargin=*, labelsep=0.5em]
    \item In a preceding offline vulnerability knowledge base construction, Vul-RAG extracts multi-dimensional knowledge from historical Common Vulnerabilities and Exposures (CVEs) and their corresponding fixes. Specifically, each knowledge item consists of three dimensions: (i) \textit{Functional Semantics}, which summarize the high-level functionality and detailed behavior of the vulnerable code; (ii) \textit{Vulnerability Causes}, which describe the root cause of the vulnerability by comparing vulnerable and patched code; and (iii) \textit{Fixing Solutions}, which summarize how the vulnerability is fixed in the patch. Vul-RAG abstracts method invocations, variable names, and types to create a representation that is less dependent on code specifics. 
    
    \item During online detection, Vul-RAG first prompts an LLM to extract the code's functional semantics. Vul-RAG then uses the code, its abstract purpose, and detailed behavior to retrieve the top-n relevant items that have high functional similarity with the input code. Retrieval is performed using probabilistic ranking (BM25) followed by a re-ranking. 
    
    \item Vul-RAG performs iterative, knowledge-augmented vulnerability detection. Instead of injecting all retrieved knowledge items into a single detection prompt, the LLM sequentially reasons over each knowledge item via (i)~a prompt for finding vulnerability causes, which checks whether the input code matches the vulnerability cause, and (ii)~a prompt for finding fixing solutions, which checks whether the code applies the corresponding fixing solution. 
    The process terminates once the code is identified as vulnerable or all items have been considered, in which case the code is classified as non-vulnerable. 
\end{enumerate}

In this study, we focus on the online retrieval and detection stages ({\protect\circled{2} and \protect\circled{3}} in Figure~\ref{fig:vul-rag}). 
To ensure controlled comparison and isolate the impact of the selected models, we keep the vulnerability knowledge base generated and provided by the authors of Vul-RAG.


\section{Experimental Setup}
\label{sec:experimental_setup}
This study aims to assess the reproducibility and replicability of Vul-RAG.
Specifically, we first examine whether the reported results can be reproduced when using the reported open-weight baseline models in a local deployment setting. 
The Vul-RAG study notes that the generalization of its findings to other LLMs requires further investigation~\cite{du2025vulragenhancingllmbasedvulnerability}. 
Motivated by this limitation, we systematically evaluate how Vul-RAG performs when replacing the reported baseline models with more recent LLMs and models with different capabilities, including code-specialized, general-purpose, and reasoning models of various parameter sizes.
We design the experiments to answer the following research questions (RQs):

\begin{enumerate}[label={\small\textbf{RQ\arabic*}}, leftmargin=4em, labelsep=0.75em]
\setlength{\itemsep}{0.1em}
    \item To what extent are the reported Vul-RAG results reproducible? 
    \item How does Vul-RAG perform with newer generation LLMs?
    \item How does model specialization (code-specialized vs. general-purpose) influence Vul-RAG performance?
    \item Do reasoning capabilities improve the performance of Vul-RAG?
    \item How does model parameter scale affect the performance of Vul-RAG?
\end{enumerate}


\subsection{Benchmark Dataset}
We reuse the benchmark dataset introduced in the Vul-RAG study.
The dataset consists of pairs of vulnerable and patched functions extracted from real-world CVEs reported for the Linux kernel. 
Each pair is enriched with Common Weakness Enumeration (CWE) identifiers and descriptions.
The dataset comprises 2,903 function pairs covering the ten most prevalent CWEs. 
It is randomly split in a 1:4 ratio into a test and training set.
The test set contains 586 pairs and serves as the evaluation benchmark (PairVul), while the remaining 2,317 pairs form the training set used to construct the knowledge base. 

The most prevalent vulnerability types in the datasets are shown in Table~\ref{tab:cwe_view}.
We categorize the CWEs according to the CWE-1000 Research View~\cite{cwe1000}, which groups weaknesses into ten high-level research pillars. This categorization enables analysis of Vul-RAG performance across vulnerability classes rather than individual CWEs. The covered CWEs span four of ten pillars, i.e., \textit{Improper Access Control}, \textit{Improper Control of a Resource Through its Lifetime}, \textit{Improper Check or Handling of Exceptional Conditions}, and \textit{Improper Neutralization}. The high coverage of resource-related vulnerabilities is consistent with the Linux kernel setting and its implementation in C, where handling of memory and system resources is a common source of vulnerabilities.

\begin{table}[t!]
\centering
\fontsize{8pt}{8pt}\selectfont
\setlength{\tabcolsep}{3pt}
\caption{Number of vulnerable-patched code pairs per CWE in PairVul and the knowledge base (KB), grouped by CWE-1000 research view pillars~\cite{cwe1000}: \textit{Improper Access Control} (Access), \textit{Improper Control of a Resource Through its Lifetime} (Resource), \textit{Improper Check or Handling of Exceptional Conditions} (Exceptional), and \textit{Improper Neutralization} (Neutralization).}
\label{tab:cwe_view}
    \begin{tabular}{>{\arraybackslash}p{1.6cm}cccccccccc}
    \toprule
    \multirow[b]{2}{*}{\raisebox{0.15\height}{\diagbox[width=1.8cm]{\textbf{Set}}{\textit{CWE}}}} & \multicolumn{1}{c}{\textbf{Access}} & \multicolumn{7}{c}{\textbf{Resource}} & \multicolumn{1}{c}{\textbf{Exceptional}} & \multicolumn{1}{c}{\textbf{Neutralization}} \\ \cmidrule{2-2} \cmidrule(l){3-9} \cmidrule(l){10-10} \cmidrule(l){11-11}
    & \textit{264} 
    & \textit{119} 
    & \textit{125} 
    & \textit{200} 
    & \textit{362} 
    & \textit{401} 
    & \textit{416} 
    & \textit{787} 
    & \textit{476} 
    & \textit{20} \\ \midrule
    
    PairVul 
    & 31 & 44 & 35 & 39 & 81 & 26 & 166 & 47 & 71 & 46 \\
    KB 
    & 120 & 173 & 140 & 152 & 320 & 101 & 660 & 187 & 281 & 182 \\
    \bottomrule
    \end{tabular}

{\fontsize{6pt}{6pt}\selectfont
\vspace{0.5em}
\raggedright
\hspace{1.25em} \textit{CWE-264}: Permissions, Privileges, and Access Controls \\
\hspace{1.25em} \textit{CWE-119}: Improper Restriction of Operations within the Bounds of a Memory Buffer \\
\hspace{1.25em} \textit{CWE-125}: Out-of-bounds Read \\
\hspace{1.25em} \textit{CWE-200}: Exposure of Sensitive Information to an Unauthorized Actor \\
\hspace{1.25em} \textit{CWE-362}: Concurrent Execution using Shared Resource with Improper Synchronization \\
\hspace{1.25em} \textit{CWE-401}: Missing Release of Memory after Effective Lifetime \\
\hspace{1.25em} \textit{CWE-416}: Use After Free \\
\hspace{1.25em} \textit{CWE-787}: Out-of-bounds Write \\
\hspace{1.25em} \textit{CWE-476}: Null Pointer Dereference \\
\hspace{1.25em} \textit{CWE-20}: Improper Input Validation \\ \vspace{-1.25em}
}
\end{table}


\subsection{Studied LLMs}
\label{sec:studied_llms}
\textbf{Baseline Models}. The Vul-RAG evaluation considered four state-of-the-art LLMs that have been widely used for vulnerability detection~\cite{du2025vulragenhancingllmbasedvulnerability}: two proprietary models (GPT-4o and Claude Sonnet 3.5) and two open-weight models (DeepSeek-Coder-V2-Instruct and Qwen2.5-Coder-32B-Instruct). 
While proprietary models are subject to version deprecation and opaque updates (rendering long-term reproducibility difficult), open-weight models remain available, ensuring that experimental results can be independently verified and compared even as the model ecosystem evolves.
\smallskip

\textbf{Additional Models}. To investigate the impact of different models, we evaluate a broader set of open-weight LLMs.
During preliminary experiments, we tested open-weights models of various parameter sizes across the Qwen, DeepSeek, Llama, and OpenAI model families, cf. online material. 
However, some of these models frequently produced outputs that did not consistently follow the given instruction format, i.e., missing \texttt{<result>YES</result>} or \texttt{<result>NO</result>} tags in \textit{\textbf{*P2}} (cf. Figure~\ref{fig:vul-rag}), and, thus, could not be reliably parsed for evaluation. 
To ensure a stable evaluation pipeline, we restrict the final set of models to those that processed all benchmark items successfully, as summarized in Table~\ref{tab:llms}. 

\begin{table}[t!]
\fontsize{8pt}{8pt}\selectfont
\setlength{\tabcolsep}{3pt}
\vspace{-1em}
\caption{Studied LLMs grouped by experiment. Full HuggingFace model identifier, parameter count, maximum context length, and knowledge cutoff.}
\label{tab:llms}
\centering
\begin{tabular}{ p{6.7cm} >{\raggedleft\arraybackslash}b{1.45cm} >{\raggedleft\arraybackslash}b{1.5cm} >{\raggedleft\arraybackslash}b{1.7cm}} \toprule
    \textbf{Model Identifier} & \textbf{Parameter} [B] & \textbf{Context} [Token] & \textbf{Cutoff Date} \\ \midrule
    
    \rowcolor{gray!10} \multicolumn{4}{l}{\textit{RQ1 Baseline models}} \\ 
     \hspace{1.mm} \textbf{deepseek-ai/DeepSeek-Coder-V2-Instruct}~\cite{deepseekai2024deepseekcoderv2breakingbarrierclosedsource} & 236 & 131,072 & Nov 2023 \\
    \hspace{1.mm} \textbf{Qwen/Qwen2.5-Coder-32B-Instruct}~\cite{hui2024qwen25codertechnicalreport} & 32.5 & 131,072 & Feb 2024 \\
    \midrule
    
    \rowcolor{gray!10} \multicolumn{4}{l}{\textit{RQ2 Newer generation of baseline models}} \\ 
    \hspace{1.mm} Qwen/Qwen3-Coder-30B-A3B-Instruct~\cite{yang2025qwen3technicalreport} & 30.5 & 262,144 & -- \\ 
    \midrule
    
    \rowcolor{gray!10} \multicolumn{4}{l}{\textit{RQ3 General-purpose models}} \\ 
    \hspace{1.mm} Qwen/Qwen2.5-32B-Instruct~\cite{qwen2.5} & 32.5 & 131,072 & -- \\
    \midrule
    
    \rowcolor{gray!10} \multicolumn{4}{l}{\textit{RQ4 Reasoning models}} \\ 
    \hspace{1.mm} deepseek-ai/DeepSeek-R1-Distill-Qwen-32B~\cite{guo2025deepseekr1} & 32.5 & 131,072 & Jul 2024 \\
    \hspace{1.mm} Qwen/QwQ-32B~\cite{qwq32b} & 32.5 & 131,072 & Nov 2024 \\
    \midrule
    
    \rowcolor{gray!10} \multicolumn{4}{l}{\textit{RQ5 Varying model parameter scale}} \\ 
    \hspace{1.mm} deepseek-ai/DeepSeek-R1-Distill-Llama-8B~\cite{guo2025deepseekr1} & 8 & 131,072 & Jul 2024 \\
    \hspace{1.mm} deepseek-ai/DeepSeek-R1-Distill-Qwen-14B~\cite{guo2025deepseekr1} & 14 & 131,072 & Jul 2024 \\
    \hspace{1.mm} Qwen/Qwen2.5-Coder-3B-Instruct~\cite{hui2024qwen25codertechnicalreport} & 3.09 & 32,768 & Feb 2024 \\
    \hspace{1.mm} Qwen/Qwen2.5-Coder-7B-Instruct~\cite{hui2024qwen25codertechnicalreport} & 7.61 & 131,072 & Feb 2024 \\
    \hspace{1.mm} Qwen/Qwen2.5-Coder-14B-Instruct~\cite{hui2024qwen25codertechnicalreport} & 14.7 & 131,072 & Feb 2024 \\
    \hspace{1.mm} Qwen/Qwen3-4B~\cite{yang2025qwen3technicalreport} & 4 & 32,768 & -- \\
    \hspace{1.mm} Qwen/Qwen3-8B~\cite{yang2025qwen3technicalreport} & 8.2 & 32,768 & -- \\
    \hspace{1.mm} Qwen/Qwen3-14B~\cite{yang2025qwen3technicalreport} & 14.8 & 32,768 & -- \\
   \hspace{1.mm} Qwen/Qwen3-30B-A3B-Instruct-2507~\cite{yang2025qwen3technicalreport} & 30.5 & 262,144 & -- \\

\bottomrule
\end{tabular}
\end{table}


\subsection{Metrics}
A challenge in vulnerability detection is that models must reliably distinguish between vulnerable code and its corresponding patch, which often differ only in small but security-relevant details~\cite{du2025vulragenhancingllmbasedvulnerability}. 
Misclassifying vulnerable code as safe (false negatives) leaves exploitable weaknesses undetected, while incorrectly flagging safe code as vulnerable (false positives) introduces unnecessary remediation effort and may reduce developer trust in automated tools.

Vul-RAG introduces the \textit{pairwise accuracy} metric, which measures the proportion of code pairs for which both the vulnerable function and its patched counterpart are correctly classified.
In addition, the framework introduces \textit{balanced recall} and \textit{balanced precision} (see Equations (1) and (2), respectively) to capture the model's ability to detect vulnerabilities and avoid false alarms. 
Balanced recall reflects the reliability of identifying vulnerable code, while balanced precision measures the accuracy of predicted vulnerabilities relative to actual weaknesses.
In the Vul-RAG study, the best performance was achieved using GPT-4o, with a pairwise accuracy of 0.32, a balanced recall of 0.58, and a balanced precision of 0.63.
To put these results in perspective, a random baseline would achieve a recall and precision of 0.50, and a pairwise accuracy of 0.25.
These results highlight the complexity of the task even for state-of-the-art models.

\begin{equation}
\fontsize{9pt}{9pt}\selectfont
    \mathrm{Balanced\ Recall} =
    \left(\frac{\#True_{\mathrm{vul}}}{\#Total_{\mathrm{vul}}} + \frac{\#True_{\mathrm{nvul}}}{\#Total_{\mathrm{nvul}}} \right) / 2 
\end{equation}

\begin{equation}
\fontsize{9pt}{9pt}\selectfont
    \mathrm{Balanced\ Precision} =
    \left(\frac{\#True_{\mathrm{vul}}}{\#Predict_{\mathrm{vul}}} + \frac{\#True_{\mathrm{nvul}}}{\#Predict_{\mathrm{nvul}}}\right) / 2
\end{equation}



\subsection{Implementation Details}
We adapted the Vul-RAG pipeline to use PyTorch and the HuggingFace Transformers library, replacing the original API-based clients with local open-weight models.
The implementation supports configurable model parameters, such as the maximum number of output tokens (\texttt{max\_new\_tokens=4096}) and temperature, which regulates the randomness of the output by scaling the probability distribution of the next-token prediction; a value close to zero ensures more deterministic and consistent responses.
For models without built-in prompt templates, we implemented a fallback formatting routine to standardize inputs. 
All other components, including source code, input data, prompts, and default Vul-RAG hyperparameter (e.g., \texttt{temperature=0.01}, \texttt{max\_knowledge=3}, \texttt{retry\_times=5}), remain unchanged.

All experiments were conducted on a high-performance computing cluster. Experiments were executed on NVIDIA L40S 48GB (for experiments with models with up to 16B parameters) or up to eight NVIDIA H100 80GB SXM5 GPUs.


\section{Results}
\label{sec:results}
In this section, we present and discuss the experimental results with respect to the research questions RQ1-RQ5 defined in Section~\ref{sec:experimental_setup}.


\subsection{RQ1 Reproducibility of Vul-RAG}
\label{sec:rq1}
To assess the reproducibility of Vul-RAG, we evaluate the framework using its reported open-weight baseline models: DeepSeek-Coder-V2-Instruct and Qwen2.5-Coder-32B-Instruct.
Following best practices for LLM reproducibility~\cite{sallouBreakingSilenceThreats2024}, we conducted three independent runs and report the mean performance.

\sisetup{
    separate-uncertainty = true, 
    table-align-uncertainty = true,
    detect-weight = true 
}
\begin{table}[b!]
\centering
\fontsize{8pt}{8pt}\selectfont
\setlength{\tabcolsep}{4pt}
\caption{Vul-RAG results reported (*) and reproduced (r). We report the mean $\pm$ standard deviation for the Qwen baseline. The DeepSeek baseline processed only 565/586 pairs; thus, the results are not representative and we report only a single run.}
\label{tab:rq1}
    \begin{tabular}{l c *{3}{S[table-format=1.2(2), table-space-text-pre=\worse]}}
    \toprule
    \textbf{LLM} & & {\textbf{Pair. Accuracy}} & {\textbf{Recall}} & {\textbf{Precision}} \\ \midrule
    
    \multirow{2}{*}{DeepSeek-Coder-V2} &
          * & 0.30 & 0.61 & 0.62 \\
        & r & \worse 0.22 & \worse 0.52 & \worse 0.53 \\ \cmidrule{2-5}
    \multirow{2}{*}{Qwen2.5-Coder-32B} &
          * & 0.26 & 0.59 & 0.61 \\ 
        & r & 0.26(1) & 0.59(1) & \worse 0.59(1) \\
    \bottomrule
    \end{tabular}
\end{table}
\begin{table}[bh!]
\centering
\fontsize{8pt}{8pt}\selectfont
\setlength{\tabcolsep}{2.15pt}
\caption{Mean CWE-specific pairwise accuracy for the Vul-RAG results reported (*) and reproduced (r). CWEs grouped by CWE-1000 research view pillars, cf. Table~\ref{tab:cwe_view}. Model names abbreviated.}
\label{tab:rq1_cwe}
\resizebox{1.01\textwidth}{!}{%
\begin{tabular}{lcrrrrrrrrrr}
\toprule
\multirow[b]{2}{*}{\raisebox{0.15\height}{\diagbox[width=2.65cm]{\textbf{LLM}}{\textit{CWE}}}} & & \multicolumn{1}{c}{\textbf{Access}} & \multicolumn{7}{c}{\textbf{Resource}} & \multicolumn{1}{c}{\textbf{Excep.}} & \multicolumn{1}{c}{\textbf{Neut.}} \\ 
\cmidrule{3-3} \cmidrule(l){4-10} \cmidrule(l){11-11} \cmidrule(l){12-12} %
    &
    & \multicolumn{1}{r}{\textit{264}}
    & \multicolumn{1}{r}{\textit{119}} 
    & \multicolumn{1}{r}{\textit{125}} 
    & \multicolumn{1}{r}{\textit{200}} 
    & \multicolumn{1}{r}{\textit{362}} 
    & \multicolumn{1}{r}{\textit{401}} 
    & \multicolumn{1}{r}{\textit{416}} 
    & \multicolumn{1}{r}{\textit{787}}
    & \multicolumn{1}{r}{\textit{476}}
    & \multicolumn{1}{r}{\textit{20}} \\ \midrule

\multirow{2}{*}{DeepSeek-Coder-V2} & * & 0.35 & 0.30 & 0.26 & 0.38 & 0.25 & 0.54 & 0.29 & 0.32 & 0.27 & 0.26 \\
     & r & \worse 0.23 & 0.30 & \worse 0.21 & \worse 0.14 & \worse 0.18 & \worse 0.29 & \worse 0.21 & \worse 0.16 & \worse 0.25 & \better 0.27 \\ \cmidrule{2-12} 
\multirow{2}{*}{Qwen2.5-Coder-32B} & * & 0.42 & 0.23 & 0.43 & 0.28 & 0.22 & 0.27 & 0.24 & 0.21 & 0.24 & 0.26 \\
     & r & \better 0.49 & \worse 0.22 & \worse 0.30 & \worse 0.26 & \better 0.25 & \worse 0.26 & \better 0.26 & \better 0.28 & \worse 0.21 & \worse 0.20 \\
\bottomrule
\end{tabular}
}%
\end{table}

Table~\ref{tab:rq1} compares the reproduced results with the results reported in the Vul-RAG study.
For Qwen2.5-Coder-32B-Instruct, the results confirm the reported performance: pairwise accuracy remains identical, and deviations in balanced recall and precision are minor (at most two percentage points, which is negligible considering the reported standard deviation).
Examining the CWE-specific pairwise accuracy in Table~\ref{tab:rq1_cwe}, we observe higher variance, e.g., in \textit{CWE-125}, we see a difference of 13 percentage points. 
However, these discrepancies are likely attributable to output variability.

\begin{figure}[b!]
    \centering
    \includegraphics[width=\linewidth,trim={0 1em 0 0},clip]{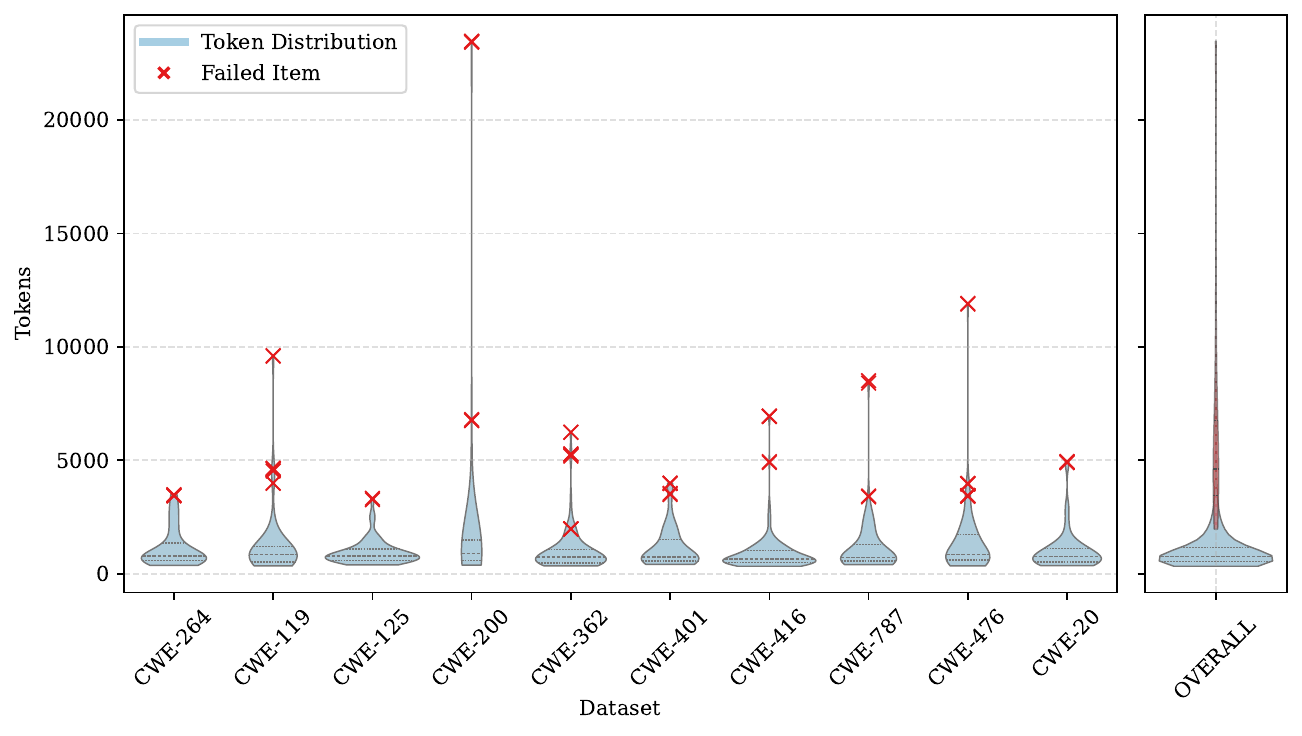}
    \caption{Distribution of tokens for the detection prompts across the PairVul CWE subsets, calculated using the DeepSeek-Coder-V2-Instruct tokenizer. Red crosses indicate dataset items where the model failed to generate a detection output. The right side shows the distribution of input code lengths of PairVul in tokens. 
    }
    \label{fig:token_distributions}
\end{figure}

For DeepSeek-Coder-V2-Instruct, we observe a performance drop across all metrics. 
DeepSeek-Coder-V2-Instruct in BF16 format requires 8$\times$80GB GPUs for inference according to its HuggingFace model card.
Despite executing on eight NVIDIA H100 GPUs, which provide the required resources, the model successfully processed only 565 of 586 pairs. 
The remaining 21 pairs repeatedly failed due to CUDA out-of-memory (OOM) errors, primarily during the early retrieval phases, e.g., extraction of function purpose and functionality (cf. \textit{\textbf{*P1}} in Figure~\ref{fig:vul-rag}).
As visualized in Figure~\ref{fig:token_distributions}, these failures affected all CWE types and were particularly pronounced for long codes. In these cases, the quadratic scaling of the self-attention mechanism exhausts available GPU memory.
Due to incomplete processing, Table~\ref{tab:rq1} and Table~\ref{tab:rq1_cwe} report only a single run for DeepSeek-Coder-V2-Instruct.
The Vul-RAG framework excludes failed items from the evaluation.
While excluding these longer and likely more complex items reflects the model’s capability on the subset it could successfully process, the results on the remaining 565 pairs still show a substantial deviation from the reported performance, with a drop of eight percentage points (0.08) in pairwise accuracy and a drop of 9 percentage points (0.09) in both recall and precision.

\answerbox{\textbf{RQ1} To what extent are the reported Vul-RAG results reproducible?}{
We were able to reproduce the results for the Qwen2.5-Coder-32B-Instruct baseline, with only minor deviations in precision. 
DeepSeek-Coder-V2-Instruct exhibits a noticeable performance drop and fails to process all input pairs due to GPU memory limitations. Importantly, failed items are excluded from the evaluation; observed discrepancies persist even on the successfully processed subset.}



\begin{table}[b!]
\centering
\fontsize{8pt}{8pt}\selectfont
\setlength{\tabcolsep}{3pt}
\caption{Vul-RAG results reported (*) and reproduced (RQ1-RQ5) using open-weight models. We highlight the \colorbox{colorFst}{\textbf{first}}, \colorbox{colorSnd}{second}, and \colorbox{colorTrd}{third} best reproduced results across models per metric; incomplete item coverage is marked in red. Model names abbreviated.}
\label{tab:rq2+}
    \begin{tabular}{llcccr}
    \toprule
    \textbf{RQ} & \textbf{LLM} & \textbf{Pair. Accuracy} & \textbf{Recall} & \textbf{Precision} & \textbf{Pairs} \\ \midrule

    \rowcolor{gray!10} \multirow{2}{*}{*} 
    & DeepSeek-Coder-V2 & 0.30 & 0.61 &  0.62 &  \\ 
    \rowcolor{gray!10} & Qwen2.5-Coder-32B & 0.26 & 0.59 & 0.61 & \\ \midrule
    
    \multirow{2}{*}{RQ1} 
    & DeepSeek-Coder-V2 & 0.22 & 0.52 &  0.53 & \textcolor{red}{565}/586 \\ 
    & Qwen2.5-Coder-32B &  0.26 & \nd 0.59 & \rd 0.59 & 586/586 \\ \midrule

    \multirow{1}{*}{RQ2} 
    & Qwen3-Coder-30B & \nd 0.28 & \nd 0.59 & \rd 0.59 & 586/586 \\ \midrule

    \multirow{1}{*}{RQ3} 
    & Qwen2.5-32B & 0.25 & \rd 0.58 & \rd 0.59 & 586/586 \\ \midrule

    \multirow{2}{*}{RQ4} 
    & DeepSeek-R1-32B & \fst 0.29 & 0.57 & 0.58 & 586/586 \\ 
    & QwQ-32B & \fst 0.29 & \fst 0.60 & \fst 0.61 & 586/586 \\ \midrule

    \multirow{9}{*}{RQ5} 
    & DeepSeek-R1-8B & \fst 0.29 & 0.57 & 0.58 & 586/586 \\
    & DeepSeek-R1-14B & \rd 0.27 & 0.56 & 0.57 & 586/586 \\
    & Qwen2.5-Coder-3B & 0.24 & 0.54 & 0.54 & 586/586 \\
    & Qwen2.5-Coder-7B & 0.08 & 0.51 & 0.52 & 586/586 \\
    & Qwen2.5-Coder-14B & 0.24 & 0.57 & 0.57 & 586/586 \\
    & Qwen3-4B & 0.25 & 0.57 & 0.58 & 586/586 \\
    & Qwen3-8B & 0.24 & 0.57 & 0.58 & 586/586 \\
    & Qwen3-14B & 0.22 & \rd 0.58 & \fst 0.61 & 586/586 \\
    & Qwen3-30B & 0.25 & \rd 0.58 & \nd 0.60 & 586/586 \\ 
    \bottomrule
    \end{tabular}
\end{table}

\subsection{RQ2 Performance with Newer Generation Models}
To assess whether improvements in models translate to enhanced vulnerability detection performance, we evaluate Vul-RAG using the newer generation of the Qwen baseline model, namely Qwen3-Coder-30B-A3B-Instruct. 
While other newer open-weight models such as DeepSeek-V2.5 and Qwen3-Coder-Next exist, their high parameter counts (236B and 80B, respectively) make them impractical for on-premise deployment; thus, we focus on the 30B-parameter model.

The results are shown in Table~\ref{tab:rq2+} row RQ2.
Qwen3-Coder-30B-A3B-Instruct achieves a modest increase of two percentage points in pairwise accuracy (0.28 vs. 0.26), while maintaining balanced recall and precision. 
However, this enhancement remains limited, suggesting that while incremental model updates might provide slight benefits in performance, they do not necessarily overcome the challenges inherent in the detection task.

\answerbox{\textbf{RQ2} How does Vul-RAG perform with newer generation LLMs?}{Incremental model upgrades yield only modest performance gains, with Qwen3-Coder-30B providing an increase of only two percentage points in pairwise accuracy over the Qwen2.5-Coder-32B baseline.}


\subsection{RQ3 Code-Specialized vs. General-Purpose LLMs}
To investigate whether strong natural language understanding benefits abstraction-based vulnerability detection, we evaluate Vul-RAG using the general-purpose counterpart of the Qwen2.5-Coder baseline, namely Qwen2.5-32B-Instruct.

The results are shown in Table~\ref{tab:rq2+} row RQ3. 
We observe only a marginal performance decrease of one percentage point in pairwise accuracy and recall compared to the code-specialized baseline.
By abstracting code and vulnerabilities into high-level semantic descriptions, Vul-RAG reframes a specialized programming task as a logical reasoning task. 
The small performance difference suggests that this abstraction can compensate for the absence of explicit code specialization and that, for RAG-based systems relying on high-level knowledge, natural language understanding capabilities may be as critical as domain-specific exposure.

\answerbox{\textbf{RQ3} How does model specialization influence Vul-RAG performance?}{The general-purpose model Qwen2.5-32B shows only a minimal performance drop of one percentage point compared to the code-specialized baseline. The results suggest that Vul-RAG’s abstraction allows general-purpose models to compensate for a lack of code-specific training.}


\subsection{RQ4 Impact of Reasoning Capabilities}
To evaluate whether explicit reasoning capabilities improve prediction, we analyze reasoning LLMs from both baseline model families, i.e., QwQ-32B and a distilled variant of the DeepSeek-R1 model (DeepSeek-R1-Distill-Qwen-32B), transferring reasoning patterns into smaller models.

\begin{figure}[b!]
    \centering
    \vspace{-.75em}
    \includegraphics[width=0.75\linewidth]{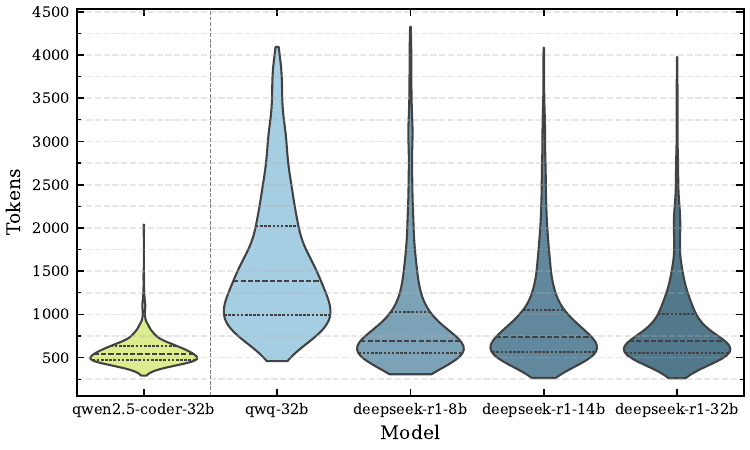}
    \vspace{-1.25em}
    \caption{Comparison of output lengths in tokens between baseline and reasoning models for \texttt{max\_new\_tokens=4096}. Reasoning models exhibit higher token counts and longer tails, reflecting the generation of reasoning steps. 
    }
    \label{fig:reasoning_output_dist}
\end{figure}
\begin{figure}[b!]
    \centering
    \includegraphics[width=1\linewidth]{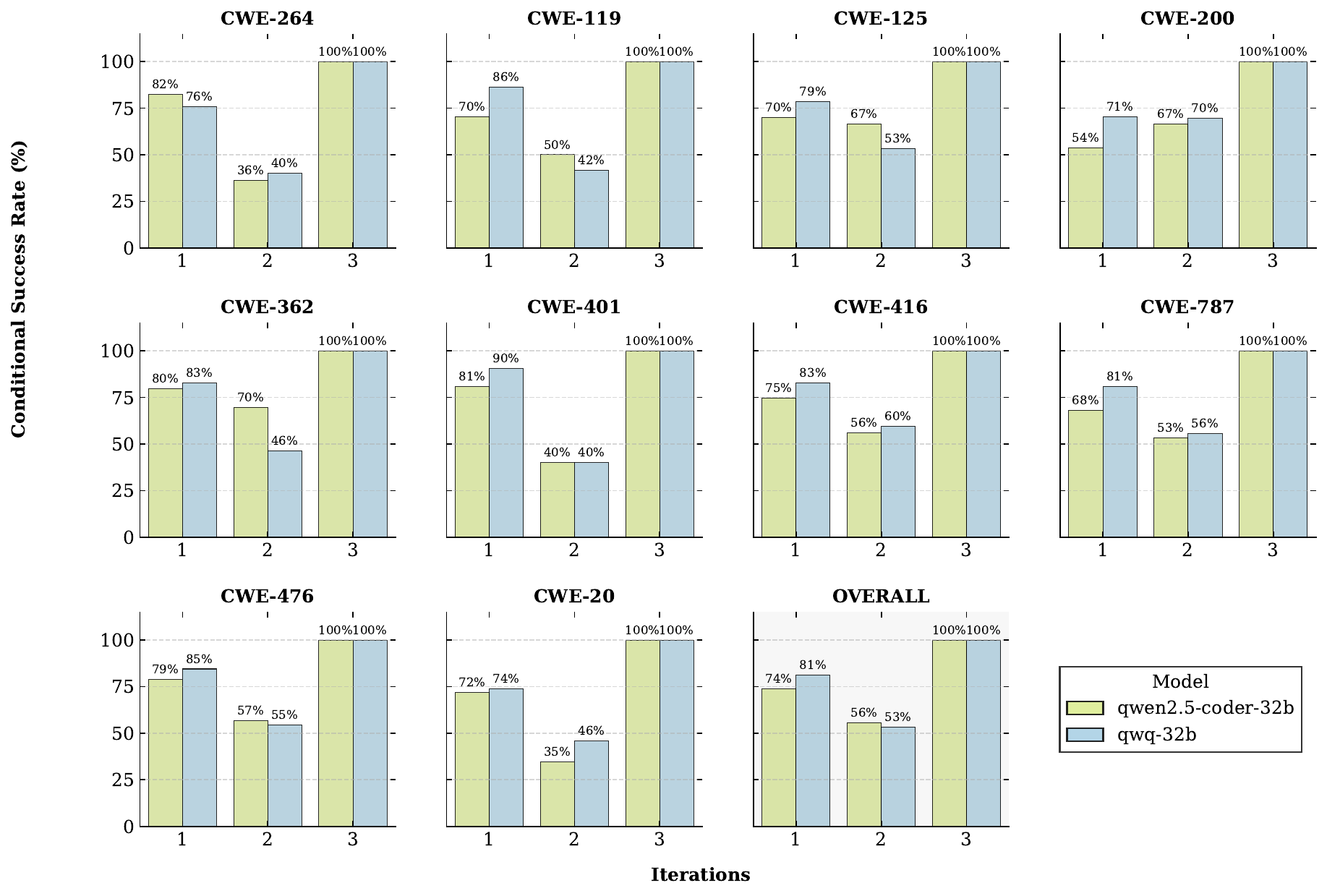}
    \vspace{-2.25em}
    \caption{Conditional success rates for reasoning vs. baseline model. Values represent the proportion of samples classified at a specific iteration, provided no decision was reached earlier. Vul-RAG iterates through up to three items before enforcing a final decision.
    }
    \label{fig:iterations}
    \vspace{-1em}
\end{figure}

The results are shown in Table~\ref{tab:rq2+} row RQ4. 
The considered reasoning models consistently outperform the baseline (RQ1), achieving the highest pairwise accuracy of 0.29, and nearly matching the reported DeepSeek-Coder-V2 baseline of 0.30. 
However, this performance gain entails a trade-off in computational overhead.
As shown in Figure~\ref{fig:reasoning_output_dist}, reasoning models generate substantially more tokens than the non-reasoning baseline.
At equal parameter scale, QwQ-32B increases the median output length from 540 to 1385 tokens (+2.6$\times$) and the IQR from 166 to 1035 tokens (+6.2$\times$) compared to Qwen2.5-Coder-32B-Instruct. 
The heavy-tailed output distribution indicates that reasoning models do not merely generate longer outputs but adopt more diverse generation behaviors.
Comparing different model families, QwQ shows extensive reasoning traces while DeepSeek-R1 produces more compact outputs. 

Further analysis of the iterative retrieval process in Figure~\ref{fig:iterations} reveals that reasoning models exhibit different conditional success rates (i.e., the proportion of codes that are classified at a specific iteration step).
The slight advantage in the first iteration suggests that they are more effective when limited contextual information is available. 
However, if no decision is reached in the first iteration, subsequent iterations provide only limited additional benefit, suggesting that the advantage of reasoning primarily lies in early-stage decision-making.

\answerbox{\textbf{RQ4} Do reasoning capabilities improve the performance of Vul-RAG?}{
Reasoning models achieve the highest pairwise accuracy but incur higher inference costs due to increased token generation.
Their advantage in iterative detection is most pronounced in the first iteration, where they more effectively translate retrieved knowledge into a correct prediction.}


\subsection{RQ5 Impact of Model Parameter Scale}
To assess the impact of model parameter scale on performance, we evaluate Vul-RAG across models from RQ1-RQ4 with parameters ranging from 3B to 32B.

The results are shown in Table~\ref{tab:rq2+} row RQ5. 
Overall, we do not observe a consistent relationship between model scale and detection performance. 
Several smaller models achieve a performance comparable to or exceeding that of larger models. 
For example, for the Qwen2.5-Coder baseline model (RQ1), performance does not drop significantly when using smaller variants, i.e., with 3B and 14B parameters. 
While the 7B variant struggled with correctly identifying pairs (biased towards predicting non-vulnerable), the 3B and 14B variants both achieve a pairwise accuracy of 0.24, only two percentage points below the baseline.
Similarly, within the Qwen3 model family (RQ2, RQ3), performance remains largely stable across scales:
The Qwen3-4B model achieves a pairwise accuracy of 0.25, matching both Qwen3-30B and Qwen2.5-32B, while Qwen3-14B performs worse with a pairwise accuracy of 0.22. 
Further, the DeepSeek-R1-Distill-Llama-8B reasoning model (RQ4) reaches a pairwise accuracy of 0.29, matching the best-performing 32B reasoning models, and even outperforms much larger non-reasoning models.
These results indicate that architectural and training refinements in newer model generations may have a greater impact than mere parameter count, and that reasoning capabilities can outweigh the benefits of increased parameter scale.

\answerbox{\textbf{RQ5} How does model parameter scale affect the performance of Vul-RAG?}{
Increasing model scale does not consistently improve detection performance. Smaller models (e.g., 4B–8B) achieve results comparable to larger 30B–32B models, and in some cases even outperform them. The findings suggest that model capabilities, such as reasoning, are more critical than parameter scale.}


\subsection{Discussion}
The results of this work provide several insights into the current state of RAG-based vulnerability detection and the specifics of the Vul-RAG framework regarding (i)~detection effectiveness, (ii)~knowledge representation and quality, and (iii)~practical deployment. \smallskip

\begin{table}[b!]
\centering
\fontsize{8pt}{8pt}\selectfont
\setlength{\tabcolsep}{2.15pt}
\caption{CWE-specific pairwise accuracy across evaluated models. We highlight the \colorbox{colorFst}{\textbf{first}}, \colorbox{colorSnd}{second}, \colorbox{colorTrd}{third} best and \colorbox{colorWrst}{worst} result per model. CWEs grouped by \textit{CWE-1000} research view pillars, cf. Table~\ref{tab:cwe_view}. Model names abbreviated.}
\label{tab:rq2+_cwe}
\resizebox{1.01\textwidth}{!}{%
\begin{tabular}{llcccccccccc}
\toprule
& \multirow[b]{2}{*}{\raisebox{-0.15\height}{\diagbox[width=2.65cm]{\textbf{LLM}}{\textit{CWE}}}} & \multicolumn{1}{c}{\textbf{Access}} & \multicolumn{7}{c}{\textbf{Resource}} & \multicolumn{1}{c}{\textbf{Excep.}} & \multicolumn{1}{c}{\textbf{Neut.}} \\ 
\cmidrule(l){3-3} \cmidrule(l){4-10} \cmidrule(l){11-11} \cmidrule(l){12-12} %
\textbf{RQ} &  
    & \textit{264}
    & \textit{119} 
    & \textit{125} 
    & \textit{200} 
    & \textit{362} 
    & \textit{401} 
    & \textit{416} 
    & \textit{787}
    & \textit{476}
    & \textit{20} \\ \midrule

\multirow{2}{*}{RQ1} & DeepSeek-Coder-V2 & 0.23 & \fst 0.30 & 0.21 & \wst 0.14 & 0.18 & \nd 0.29 & 0.21 & 0.16 & 0.25 & \rd 0.27 \\
& Qwen2.5-Coder-32B & \fst 0.49 & 0.22 & \nd 0.30 & 0.26 & 0.25 & 0.26 & 0.26 & \rd 0.28 & 0.21 & \wst 0.20  \\ \midrule

\multirow{1}{*}{RQ2} & Qwen3-Coder-30B & \fst 0.39 & 0.25 & \nd 0.37 & \wst 0.23 & 0.28 & \wst 0.23 & 0.25 & \rd 0.34 & 0.31 & 0.28 \\ \midrule

\multirow{1}{*}{RQ3} & Qwen2.5-32B & \fst 0.45 & \wst 0.20 & 0.23 & \nd 0.28 & 0.25 & \rd 0.27 & 0.24 & 0.23 & 0.24 & \wst 0.20 \\ \midrule

\multirow{2}{*}{RQ4} & DeepSeek-R1-32B & \fst 0.55 & \rd 0.32 & 0.31 & \nd 0.33 & \nd 0.33 & 0.23 & 0.27 & \rd 0.32 & 0.23 & \wst 0.13 \\
& QwQ-32B & \fst 0.48 & 0.27 & \rd 0.37 & 0.28 & 0.25 & \nd 0.46 & 0.32 & \wst 0.17 & 0.20 & 0.22 \\ \midrule

\multirow{9}{*}{RQ5} & DeepSeek-R1-8B & 0.29 & 0.30 & 0.26 & \nd 0.33 & 0.25 & \fst 0.42 & 0.28 & 0.30 & \rd 0.32 & \wst 0.24 \\
& DeepSeek-R1-14B & 0.23 & 0.23 & \nd 0.34 & 0.26 & \fst 0.36 & 0.27 & 0.27 & \rd 0.28 & 0.27 & \wst 0.20 \\
& Qwen2.5-Coder-3B & 0.26 & 0.20 & \wst 0.14 & \fst 0.36 & 0.22 & \nd 0.35 & 0.20 & 0.21 & 0.27 & \rd 0.33 \\
& Qwen2.5-Coder-7B & \wst 0.06 & 0.07 & \nd 0.11 & \rd 0.10 & \wst 0.06 & \fst 0.15 & 0.08 & \wst 0.06 & \rd 0.10 & 0.07 \\
& Qwen2.5-Coder-14B & 0.23 & \nd 0.27 & \fst 0.34 & 0.21 & 0.23 & 0.23 & \rd 0.24 & 0.21 & \nd 0.27 & \wst 0.17 \\ 
& Qwen3-4B & \rd 0.29 & 0.25 & \fst 0.34 & 0.23 & 0.26 & 0.27 & 0.24 & \nd 0.30 & 0.20 & \wst 0.17 \\
& Qwen3-8B & \fst 0.32 & \rd 0.30 & 0.26 & 0.23 & \wst 0.22 & \nd 0.31 & 0.23 & 0.23 & 0.23 & \wst 0.22 \\
& Qwen3-14B & \fst 0.35 & 0.20 & \nd 0.29 & 0.18 & 0.16 & \fst 0.35 & 0.22 & \rd 0.26 & 0.20 & \wst 0.13 \\
& Qwen3-30B & \fst 0.55 & \rd 0.25 & 0.23 & 0.23 & 0.20 & \nd 0.35 & 0.23 & \wst 0.19 & \rd 0.25 & 0.20 \\
\bottomrule

\end{tabular}
}%

{\fontsize{6pt}{6pt}\selectfont
\vspace{0.5em}
\raggedright
\hspace{1.25em} \textit{CWE-264}: Permissions, Privileges, and Access Controls \\
\hspace{1.25em} \textit{CWE-119}: Improper Restriction of Operations within the Bounds of a Memory Buffer \\
\hspace{1.25em} \textit{CWE-125}: Out-of-bounds Read \\
\hspace{1.25em} \textit{CWE-200}: Exposure of Sensitive Information to an Unauthorized Actor \\
\hspace{1.25em} \textit{CWE-362}: Concurrent Execution using Shared Resource with Improper Synchronization \\
\hspace{1.25em} \textit{CWE-401}: Missing Release of Memory after Effective Lifetime \\
\hspace{1.25em} \textit{CWE-416}: Use After Free \\
\hspace{1.25em} \textit{CWE-787}: Out-of-bounds Write \\
\hspace{1.25em} \textit{CWE-476}: Null Pointer Dereference \\
\hspace{1.25em} \textit{CWE-20}: Improper Input Validation \\ \vspace{-1.25em}
}
\end{table}

\textbf{Detection effectiveness.}
A primary finding across all research questions RQ1-RQ5 is the narrow performance difference despite the diversity of the models evaluated, cf. Table~\ref{tab:rq2+}.
Notably, the highest achieved overall pairwise accuracy of 0.29 is only four percentage points above the random baseline of 0.25.
Hence, while the Vul-RAG framework enhances the ability of LLMs to distinguish between vulnerable and non-vulnerable code~\cite{du2025vulragenhancingllmbasedvulnerability}, these capabilities do not scale with model advancement; for instance, Qwen3-4B achieves the same pairwise accuracy (0.25) as Qwen3-30B-A3B-Instruct-2507.
The observation of a performance plateau is also evident in the study by Antal et al.~\cite{antalEvaluatingRetrievalAugmentedGeneration2026}, where we observe a similar ceiling in the context of Java vulnerabilities. 
Despite evaluating (proprietary) frontier models with large parameter scales, e.g., DeepSeek-V3.1, GPT-5, and Llama-4-Maverick, the authors reported pairwise accuracies of 16.7\%, 23.1\%, and 32.4\% respectively. 
The persistence of this plateau across different programming languages, datasets, knowledge bases, and models reinforces the inherent difficulty of the pairwise discrimination task.

Performance also varies considerably across vulnerability types, as shown in Table~\ref{tab:rq2+_cwe}, indicating that certain vulnerabilities are inherently more difficult to capture, even with abstract knowledge representations.
Across all models, performance is consistently highest for \textit{CWE-264}, where access control vulnerabilities often follow well-defined patterns that can be captured through semantic descriptions. 
In contrast, performance is lowest for \textit{CWE-20}. Input validation logic is typically distributed across multiple code locations and depends on broader program context, making it difficult to represent through localized functional semantics.
For resource-related CWEs, performance remains inconsistent across models. These vulnerabilities often involve complex interactions or implicit assumptions about program state, which are not easily captured by static knowledge snippets. 
Notably, reasoning models (RQ4) exhibit more stable performance across CWE categories, suggesting that explicit reasoning helps mitigate some of the limitations of incomplete or imperfect retrieval. 

\smallskip


\textbf{Knowledge representation and quality.}
For reproducibility, we reused the Vul-RAG knowledge base, which was generated using GPT-3.5-turbo-0125. 
While necessary to ensure a consistent baseline, this choice raises questions about whether system performance is effectively capped by the quality of the initial knowledge extraction.
The analysis in Vul-RAG identified that false negatives often stem from non-existent or vague vulnerability descriptions within the knowledge base or from irrelevant retrieval, where the retrieved functional semantics are too different from the input function. 
Similarly, false positives are frequently triggered by mismatched fixing solutions, particularly when a vulnerability could be addressed by alternative patches not represented in the retrieved context~\cite{du2025vulragenhancingllmbasedvulnerability}.
It remains unclear whether results would improve if the knowledge base were generated by, e.g., a reasoning model.
Future research should, therefore, investigate where model capability matters most (whether in the construction of the knowledge base, the summarization of code semantics, or the final vulnerability prediction) and how to evaluate such capability regarding quality. \smallskip


\textbf{Practical deployment.}
We conducted all experiments on a high-performance computing cluster. 
Such extensive computational resources are typically unavailable in industrial development environments, where hardware is often more constrained. 
This discrepancy highlights the importance of the findings regarding model scale: 
The small performance differences between models suggest that selecting significantly larger models does not necessarily yield proportional gains. 
Notably, several smaller models achieved comparable performance. These smaller models are more attractive for on-premise deployment, as they can operate on commodity hardware while maintaining privacy.

Overall, several challenges remain regarding the real-world practicality of RAG-based systems, specifically Vul-RAG.
The pairwise accuracy ceiling of approximately 0.30 (and the resulting false positives and false negatives) suggests that Vul-RAG currently serves more as a research framework than a production tool. 
In addition, its current scope is limited to the Linux kernel and ten CWE types.
The non-deterministic nature of LLMs and the strict formatting requirements also pose a challenge for automated integration.
Further, the use of a binary output without providing verifiable reasoning makes it difficult for developers to interpret or trust the results.
Finally, further benchmarking is required to determine whether these findings generalize to other programming languages and vulnerability types.


\section{Threats to Validity}
\label{sec:threats}
Several factors may impact the validity of the findings. \smallskip

\textbf{Data leakage.} 
A primary threat in LLM-based research is data leakage, i.e., the possibility that a model has seen the evaluation samples during its training and, thus, memorized correct answers rather than performing genuine vulnerability detection.
The PairVul dataset was initially published on GitHub in April 2024. 
Given that several models used in this study were released or updated after this date (specifically QwQ-32B, variants of Qwen-2.5 and Qwen3, cf. Table~\ref{tab:llms}), we cannot rule out the possibility that the evaluation samples were included in their training data. 
The lack of transparency regarding the training data of LLMs remains an open reseacrh challenge. \smallskip

\textbf{Bias in model selection.}
Vul-RAG relies on \texttt{<result>YES</result>} or \texttt{<result>NO</result>} tags in the model response for automated parsing of the prediction. 
During preliminary experiments, we excluded several models from discussion in the RQs, as they failed to consistently adhere to this format. 
A model’s failure to follow specific output formats does not necessarily imply an inability to detect vulnerabilities. 
However, with the setup used by Vul-RAG, such models are penalized, introducing a selection bias towards models with superior instruction-following capabilities. 
Loosening the evaluation script to handle free-form text might yield different results for less-instructed models but would introduce non-deterministic parsing errors.
\smallskip

\textbf{Computational resources and deployment.}
Hardware constraints affected the evaluation of DeepSeek-Coder-V2, for which we encountered CUDA out-of-memory errors on an 8$\times$80GB H100 node. 
Consequently, the performance reported for this model is not fully representative. 
More importantly, we highlight a significant gap between research and practice: the infrastructure required to run such large models exceeds the resources available to typical mid-sized organizations for software vulnerability detection, underscoring the necessity for evaluating smaller, more resource-efficient models that can be realistically deployed in industry pipelines.


\section{Conclusion}
\label{sec:conclusion}
This work presents a reproducibility and replicability study of the Vul-RAG framework in a fully local, open-weights setting.
We find that Vul-RAG is reproducible for Qwen2.5-Coder but not for DeepSeek-Coder-V2.
Across a diverse set of LLMs, including newer model generations~(RQ2), general-purpose models~(RQ3), and reasoning models~(RQ4) with varying parameter scales~(RQ5), we observe only small performance differences.
Reasoning models achieve the highest pairwise accuracy of 0.29.
In contrast, neither increased model scale nor newer model generations yield substantial improvements.
These results consistently indicate a performance plateau in Vul-RAG and highlight the inherent difficulty of the pairwise discrimination task.
From a practical perspective, the findings show that large parameter scales are not required for competitive performance, allowing for efficient deployment with smaller models. 
Future work should focus on improving the knowledge base quality and retrieval strategies and investigate where model capability matters most in RAG-based pipelines to overcome the existing performance limitations.


\begin{credits}
\subsubsection{\ackname} 
This work has been funded by the Federal Ministry of Research, Technology and Space (BMFTR) and the state of Baden-Württemberg (Program: Forschung an HAW, Grant No. 13HAW26PX4), and in part by the German Research Foundation (DFG) under {project-ID 528745080 - FIP 68}. The authors alone are responsible for the content of this paper.
We thank the DACHS data analysis cluster, hosted at Hochschule Esslingen and co-funded by the MWK within the DFG's "Großgeräte der Länder" program, for providing the computational resources necessary for this research.

\end{credits}

\bibliographystyle{splncs04}
\bibliography{literature}

\end{document}